\documentclass[%
 reprint,
bibnotes,
 amsmath,amssymb,
 aps,
]{revtex4-1}

\usepackage{xprintlen}

\usepackage{graphicx}
\usepackage[english]{babel}
\usepackage{microtype}          
\usepackage[breaklinks=true,colorlinks=true,linkcolor=blue,urlcolor=blue,citecolor=blue]{hyperref}

\usepackage{xcolor}     
\usepackage{mathtools}
\usepackage{braket}
\usepackage{tikz}
 \usepackage{printlen}

 \usepackage{longtable}
  \usepackage{booktabs}
 \usepackage{siunitx}
 \usepackage{csquotes}

\usepackage{dcolumn}
\usepackage{bm}

\begin{document}
\bibliographystyle{apsrev}
\preprint{APS/123-QED}

\title{\textit{Ab initio} Exchange-Correlation Free Energy\\ of the Uniform Electron Gas at Warm Dense Matter Conditions}

\author{Simon Groth$^{1,\dagger}$}
 \email{groth@theo-physik.uni-kiel.de}
\author{Tobias Dornheim$^{1,\dagger}$}%
\author{Travis Sjostrom$^{2}$}
\author{Fionn D.~Malone$^{3}$}
\author{W.M.C.~Foulkes$^{3}$}
\author{Michael Bonitz$^{1}$}
\affiliation{$^\dagger$These authors contributed equally to this work.\\ $^1$Institut f\"ur Theoretische Physik und Astrophysik, Christian-Albrechts-Universit\"{a}t zu Kiel, D-24098 Kiel, Germany\\ $^2$Theoretical Division, Los Alamos National Laboratory, Los Alamos, New Mexico 87545, USA\\ 
$^{3}$Department of Physics, Imperial College London, Exhibition Road, London SW7 2AZ, UK}

\date{\today}

\begin{abstract}
In a recent Letter [T.~Dornheim \textit{et al.}, Phys.~Rev.~Lett.~\textbf{117}, 156403 (2016)], we presented the first \textit{ab initio} quantum Monte-Carlo (QMC) results of the warm dense electron gas in the thermodynamic limit. However, a complete parametrization of the exchange-correlation free energy with respect to density, temperature, and spin polarization remained out of reach due to the absence of (i) accurate QMC results below $\theta=k_\text{B}T/E_\text{F}=0.5$ and (ii) of QMC results for  spin polarizations different from the paramagnetic case. Here we overcome both remaining limitations. By closing the gap to the ground state and  by performing extensive QMC simulations for different spin polarizations, we are able to obtain the first complete \textit{ab initio} exchange-correlation free energy functional; the accuracy achieved is an unprecedented $\sim 0.3\%$. This also allows us to quantify the accuracy and systematic errors of various previous approximate functionals.
\end{abstract}

\maketitle

%

The uniform electron gas (UEG), i.e., Coulomb interacting electrons in a homogeneous positive background, is one of the seminal model systems in physics \cite{loos}. Studies of the UEG led to key insights such as Fermi liquid theory \cite{quantum_theory,quantum_theory2}, the quasi-particle picture of collective excitations \cite{pines,pines2}, and BCS theory of superconductivity \cite{bcs}. Furthermore, accurate parametrizations of its ground state 
properties \cite{vwn,perdew,pw,cdop,gori,gori2} based on \textit{ab initio} quantum Monte-Carlo (QMC) simulations \cite{gs1,gs2,ortiz,ortiz2,spink} have sparked many applications \cite{farid,pdw_map,takada} and facilitated the success of density functional theory (DFT) simulations of atoms, molecules, and real materials \cite{ks,dft_burke,dft_review}.

However, over the last decade, there has emerged an increasing interest in matter under extreme excitation or compression, such as laser-excited solids \cite{ernst} and inertial confinement fusion targets \cite{nora,schmit,hurricane3,kritcher}. Astrophysical examples such as white dwarf atmospheres and planet interiors \cite{knudson,militzer} provide further motivation. This so-called warm dense matter (WDM) regime \cite{wdm_book} is characterized by values close to unity of the Wigner-Seitz radius (quantum coupling parameter) $r_s=\overline{r}/a_\textnormal{B}$ and degeneracy parameter $\theta=k_\textnormal{B}T/E_\textnormal{F}= 2k_\textnormal{B}T/(6\pi^2n^\uparrow)^{(2/3)}$. Here $E_\textnormal{F}$ denotes the $\xi$-dependent Fermi energy, $n_\uparrow = N^\uparrow / \Omega$ the particle number per volume, $\overline{r}$ the mean interparticle distance, and $a_B$ the Bohr radius. A third parameter is the degree of spin polarization, $\xi = (N^\uparrow-N^\downarrow)/(N^\uparrow+N^\downarrow)$, where $N^\uparrow$ ($N^\downarrow$) is the number of spin-up (spin-down) electrons. An accurate theoretical description of this exotic state is most challenging since it must capture the nontrivial interplay of coupling, excitation, and quantum degeneracy effects.
Naturally, an accurate parametrization of the exchange-correlation (XC) free energy per electron, $f_\text{xc}$, of the UEG at WDM conditions is a fundamental step towards this goal as it constitutes key input for, e.g., thermal DFT \cite{mermin,holst,burke_warm}, quantum hydrodynamics \cite{manfredi,michta}, and the construction of equations of state for astrophysical objects~\cite{pot1,sauron1,sauron2}.

\begin{figure*}
\includegraphics[width=0.32\textwidth]{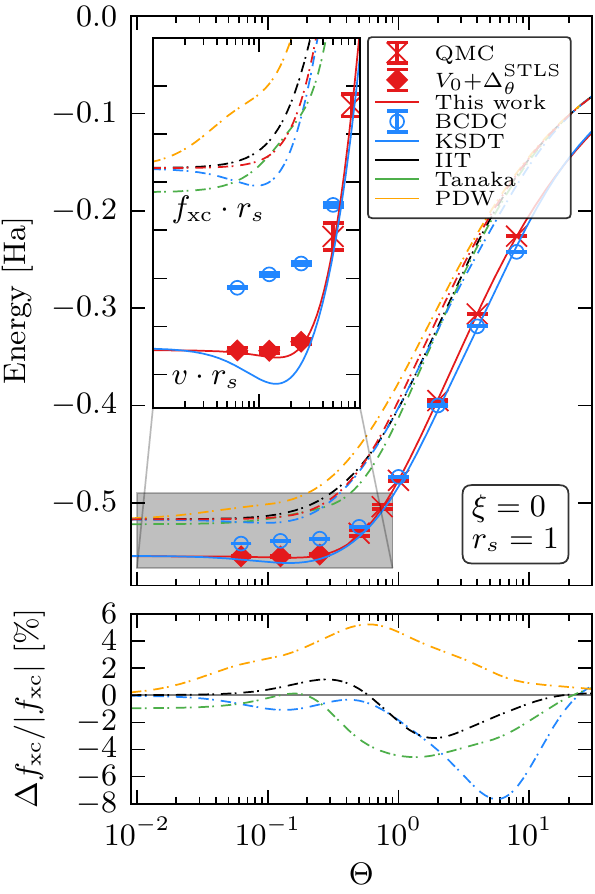}
\includegraphics[width=0.32\textwidth]{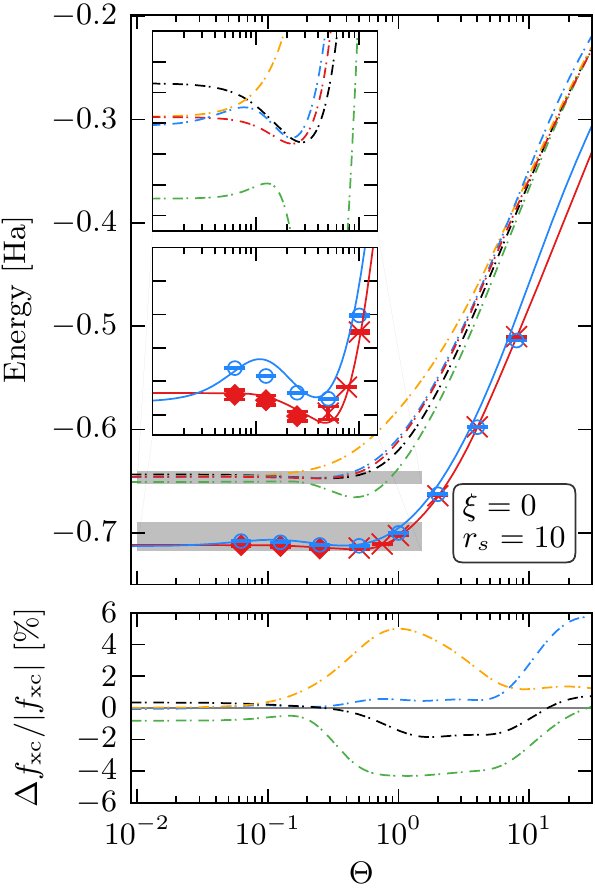}
\includegraphics[width=0.32\textwidth]{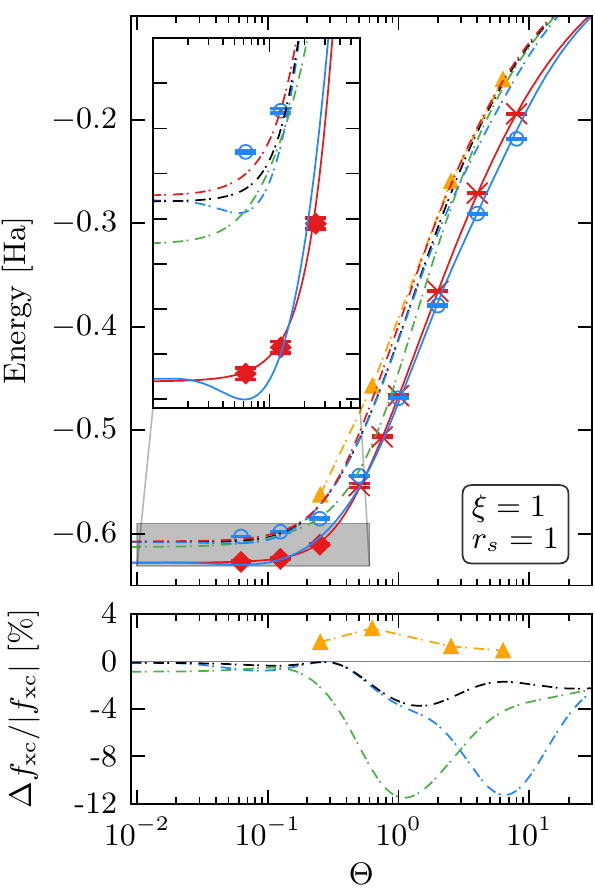}

\caption{\label{fig:kachel_mann}Temperature dependence of the XC free energy and potential energy - The top row shows $f_\text{xc}$ (dashed lines) from this work (red), KSDT (blue, \cite{ksdt}), IIT (black, \cite{tanaka_old,tanaka_new}), Tanaka (green, \cite{tanaka_hnc}), and PDW (yellow dashed line and triangles, \cite{pdw_param}), as well as the corresponding interaction energy $v$ (solid lines) from this work, KSDT, and the RPIMC results by Brown \textit{et al.}~(blue dots, \cite{bcdc}). The red rhombs correspond to ground state QMC results plus a temperature correction function $\Delta_\theta^\textnormal{STLS}$, see Eq.~(\ref{eq:correction}). The inset corresponds to a zoom into the grey box. The bottom row displays the relative deviations of the different models of $f_\text{xc}$ with respect to our new parametrization.
}
\end{figure*}

The first parametrizations of $f_\text{xc}$ were constructed based on uncontrolled approximations such as interpolations between known limits~\cite{ebeling1}, semi-empirical quantum-classical mappings \cite{pdw_map,pdw_param}, and dielectric (linear response) methods \cite{stls,stls2,tanaka_hnc,tanaka_old,tanaka_new}. 
\textit{Ab initio} QMC simulations of the UEG are severely limited by the fermion sign problem (FSP) \cite{loh,troyer}, so the pioneering results of Brown \textit{et al.}~\cite{bcdc} were based on the restricted path integral MC (RPIMC) approach, in which the nodal structure of the density matrix is assumed. However, the accuracy of the RPIMC data has recently been challenged, as they are afflicted with systematic errors exceeding $10\%$ at $r_s=1$~\cite{tim3}. Nevertheless, these data were used as input for several parametrizations of $f_\text{xc}$ \cite{stls2,bdhc,ksdt},  the most sophisticated being that of Karasiev \textit{et al.} (KSDT)~\cite{ksdt}.
Unsurprisingly, all the aforementioned models substantially deviate from each other (cf.~Fig.~\ref{fig:kachel_mann}) in the WDM regime~\cite{groth2}.

This unsatisfactory situation has sparked remarkable recent progress in the field of fermionic QMC simulations 
\cite{filinov_pre15, dubois_arxiv,tim_cpp15,tim3,dornheim,dornheim2,malone,malone2,dornheim_prl,dornheim_pop}. In particular, the combination of three complementary QMC methods-- configuration PIMC (CPIMC) \cite{tim3}, permutation blocking PIMC (PB-PIMC) \cite{dornheim,dornheim2}, and density matrix QMC (DM-QMC) \cite{malone,malone2}--allows one to avoid the FSP over a broad parameter range {\em without any nodal bias}~\cite{groth,dornheim3}. In a recent Letter \cite{dornheim_prl} we presented an improved procedure to extrapolate the QMC results to the thermodynamic limit and thereby obtained \textit{ab initio} data for the unpolarized UEG with an unprecedented accuracy of the order of $0.1\%$.
However, two remaining problems prohibited the construction of a complete parametrization of $f_\text{xc}$ with respect to $r_s$, $\theta$, and $\xi$: (i) the absence of accurate QMC results for $0<\theta<0.5$, due to the FSP, and (ii) the lack of QMC results for $\xi>0$ and, thus, of an appropriate spin interpolation function $\Phi(r_s,\theta,\xi)$. This is used, for example, in DFT calculations in the local spin density approximation, which require the evaluation of $f_{xc}$ at arbitrary values of $\xi$.

In this Letter, both remaining problems are solved. Inspired by Tanaka and Ichimaru~\cite{tanaka_old,tanaka_new} and the previously discovered~\cite{groth2} impressive accuracy of the Singwi-Tosi-Land-Sj\"olander (STLS) formalism~\cite{stls,stls2} (STLS), we bridge the gap between $\theta=0$ and $\theta=0.25$ by adding the (small) temperature dependence of STLS,
\begin{eqnarray}
\Delta_\theta^\text{STLS}(r_s,\theta,\xi) = v^\text{STLS}(r_s,\theta,\xi) - v^\text{STLS}(r_s,0,\xi),
\label{eq:correction}
\end{eqnarray}
to the known exact ground-state QMC interaction energy $v_0$~\cite{spink}. Second, we carry out extensive QMC simulations of the warm dense UEG for $\xi=1/3,0.6$, and $1$ (179 data points ranging from $0.1\leq r_s \leq 20$ and $0.5\leq \theta \leq 8$, see Table 3 in the Supplemental Material~\cite{supplement}). In combination with the results from Ref.~\cite{dornheim_prl} this allows us to construct the first complete \textit{ab initio} parametrization of the XC free energy, $f_\text{xc}(r_s,\theta,\xi)$, and to attain an unprecedented accuracy of~$\sim0.3\%$. 
The high quality of our new results is verified by various cross-checks and compared to the widely used parametrizations by Karasiev \textit{et al.}~(KSDT~\cite{ksdt}), Perrot and Dharma-wardana~(PDW~\cite{pdw_param}), Ichimaru, Iyetomi, and Tanaka~(IIT~\cite{tanaka_old,tanaka_new}), and the recent improved dielectric approach by Tanaka~\cite{tanaka_hnc}.

{\bf Parametrization of $f_\text{xc}$ for $\xi=0$ and $\xi=1$}.
Following Refs.~\cite{tanaka_old,tanaka_new} we obtain $f_\text{xc}$ from our {\em ab initio} QMC data for 
the interaction energy~$v^\xi(r_s,\theta)$ via
\begin{eqnarray}
f^\xi_\text{xc}(r_s,\theta) &=& \frac{1}{r_s^2} \int_0^{r_s} \textnormal{d}\overline{r}_s\ \overline{r}_s v^\xi(\overline{r}_s,\theta) \\
\label{eq:fit}\Rightarrow v^\xi(r_s,\theta) &=& \left. 2f^\xi_\text{xc}(r_s,\theta) + r_s \frac{ \partial f^\xi_\text{xc}(r_s,\theta) }{ \partial r_s }\right|_\theta\ .
\end{eqnarray}
We employ Pad\'{e} formulas for $f^1_\text{xc}$ and $f^0_\text{xc}$~\cite{supplement},
\begin{align}
f^\xi_\text{xc}(r_s,\theta) = 
 -\frac{1}{r_s} \frac{\omega_\xi a(\theta) + b^\xi(\theta)\sqrt{r_s}+c^\xi(\theta)r_s }{ 1+d^\xi(\theta)\sqrt{r_s}+e^\xi(\theta)r_s }\ , \label{eq:fxc-pade}
\end{align}
where $\omega_0=1$, $\omega_1=2^{1/3}$, and the functions $a$-$d$ are given in the Supplemental Material~\cite{supplement}. We fit the RHS~of Eq.~(\ref{eq:fit}) to our combined data for $v^{1,0}$. To ensure the correct ground state limit, we use the relation~\cite{ksdt}
\begin{eqnarray}
\lim_{\theta\to0} f_\text{xc}^\xi(r_s,\theta) = e^\xi_\text{xc}(r_s,0) \ 
\end{eqnarray}
to fit the zero temperature limit of our Pad\'e formula to the recent QMC results by Spink \textit{et al.}~\cite{spink}. In addition, the classical Debye-H\"uckel limit for large $\theta$ and the Hartree-Fock limit $f_\text{xc}^\textnormal{HF}(r_s,\theta)=a(\theta)/r_s \equiv a^\textnormal{HF}(\theta)/r_s$~\cite{pdw84} for $r_s\to0$ are exactly incorporated.

The new \textit{ab initio} results for $f_\text{xc}^\xi(r_s,\theta)$ are depicted in Fig.~\ref{fig:kachel_mann} (red dashed line) and compared to various approximations. While all curves exhibit a qualitatively similar behavior with respect to temperature, there appear substantial deviations for intermediate $\theta$ between $5\ldots 12\%$ (bottom row). We find that, for $\xi=0$, the IIT parametrization exhibits the smallest errors, whereas, for $\xi=1$, the PDW points are superior, although the IIT curve is of a similar quality. The recent parametrization by Tanaka (green) does not constitute an improvement compared to IIT. Finally, the KSDT curves are relatively accurate at low $\theta$, but systematically deviate for $\theta \gtrsim 0.5$, in particular towards higher density ($r_s\lesssim 4$, \cite{supplement}), with a maximum deviation of $\Delta f/f\sim10\%$. This can be traced to an inappropriate finite-size correction of the QMC data by Brown \textit{et al.}~\cite{bcdc} (BCDC), see Ref.~\cite{dornheim_prl}. The deviations are even more severe for $\xi=1$, in agreement with previous findings about the systematic bias in the RPIMC input data~\cite{groth, dornheim3} and with recent investigations~\cite{tanaka_hnc,tanaka_new} of $f_\text{xc}$ itself. Also notice the pronounced bump of $f_\text{xc}^0$ occuring for large $r_s$ and low temperature (see inset in the middle panel), which induces an unphysical negative total entropy~\cite{burke} in the KSDT fit.

Consider now the red rhombs and crosses in Fig.~\ref{fig:kachel_mann} that show our new \textit{ab initio} data for the interaction energy. We observe a smooth connection between our QMC data for $\theta\geq0.5$ (crosses) and the temperature-corrected ground state data $v_0+\Delta_\theta^\textnormal{STLS}$ [Eq.~(\ref{eq:correction})] (rhombs) in all three parts of the figure. This behavior is observed 
for all densities. The solid red line depicts the fit to $v^\xi$, cf.~Eq.~(\ref{eq:fit}). The utilized Pad\'e ansatz is an extraordinarily well suited fit function as it reproduces the input data ($v^\xi$) for $\xi=0$ ($\xi=1$) with a mean and maximum deviation of $0.12\%$ and $0.68\%$ ($0.17\%$ and $0.63\%$)~\cite{ksdt-bcdc-note}.

\begin{figure}
\includegraphics{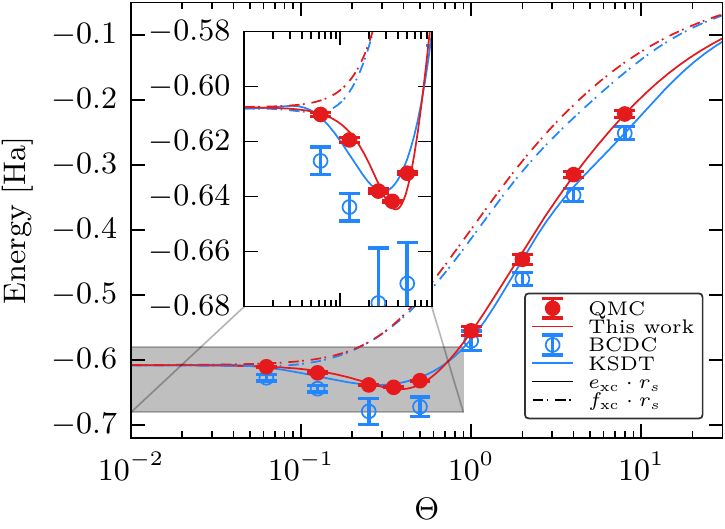}
\caption{\label{fig:best_plot_in_human_history}
Cross-check of our parametrization ($\xi=1$, $r_s=1$). The XC energy per electron (red line), as calculated from our Pade function for $f_{xc}$ (dashed line), is compared to new, independent finite-size-corrected QMC data (red dots)~\cite{supplement}. While our functional has been constructed solely using the interaction energy $v$, cf.~Eq.~(\ref{eq:fit}), the KSDT curve~\cite{ksdt} (solid blue) was fitted to the BCDC data \cite{bcdc} for $e_\text{xc}$ (blue circles).
}
\end{figure}

To further illustrate the high quality of our XC functional and to verify the applied temperature correction (\ref{eq:correction})  at low $\theta$, we
carried out extensive new QMC simulations for the exchange-correlation energy per particle, $e_\text{xc}$, for $r_s=1$ and $\xi=1$,
over the entire $\theta$-range down to $\theta=0.0625$ (see Ref.~\cite{supplement} for details). The finite-size-corrected data are compared to $e_\text{xc}$ reconstructed from our parametrization of $f_\text{xc}^\xi(r_s,\theta)$ via~\cite{ksdt}
\begin{eqnarray}
e^\xi_\text{xc}(r_s,\theta) = f_\text{xc}^\xi(r_s,\theta) - \theta \left. \frac{ \partial f_\text{xc}^\xi(r_s,\theta) }{ \partial \theta}\right|_{r_s}\ . \label{eq:exc}
\end{eqnarray}
This allows us not only to gauge the accuracy of $f_\text{xc}$ itself but also its temperature derivative, which is directly linked to the XC-entropy.
The results are presented in Fig.~\ref{fig:best_plot_in_human_history} and demonstrate excellent agreement between our parametrization (red solid line) and the independent new QMC data (red dots) over the entire range of $\theta$. Since the new data for $e_\text{xc}$ were not used for our fit this constitutes a further impressive confirmation of Eq.~(\ref{eq:correction}) and demonstrates the consistency of our parametrization. 
This is in stark contrast to previous works (see blue symbols and line)~\cite{ksdt-bcdc-note,ksdt-note-2}.

\begin{figure}
\includegraphics{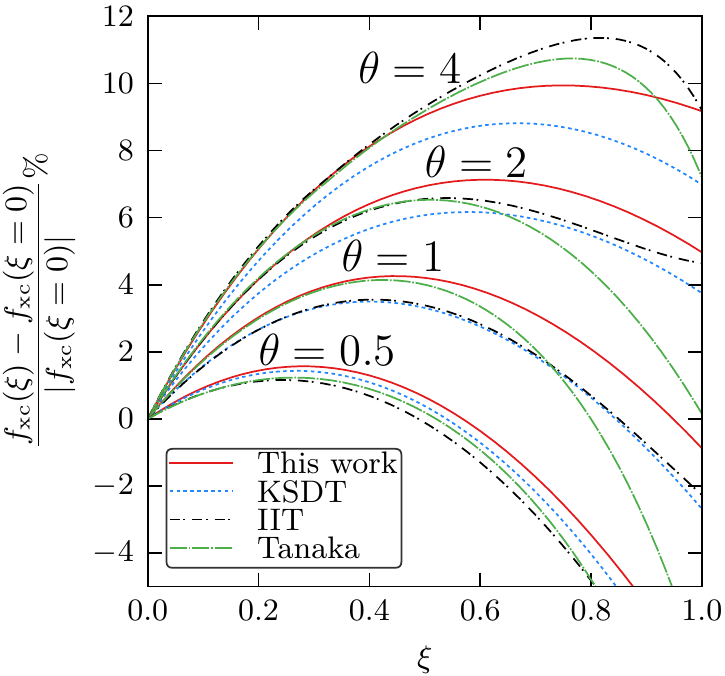}
\caption{Dependence of the XC-free energy on spin polarization at $r_s=1$. The \textit{ab initio} functional of this work (red) is compared to the parametrizations by Karasiev \textit{et al.}~\cite{ksdt} (KSDT, blue), Ichimaru \emph{et al.}~\cite{tanaka_old,tanaka_new} (IIT, black), and Tanaka  (green) \cite{tanaka_hnc}.  \label{fig:xi}
}
\end{figure}

{\bf Spin interpolation.}
To obtain an accurate parametrization of $f_\text{xc}$ at arbitrary spin polarization $0\leq \xi \leq 1$, we employ the ansatz~\cite{pdw_param}
\begin{eqnarray}
\nonumber f_\text{xc}(r_s,\theta,\xi) &=& f^0_\text{xc}(r_s,\theta^0) + \Big[f_\text{xc}^1(r_s,\theta^{0}\cdot 2^{-2/3}) \\ & & - f^0_\text{xc}(r_s,\theta^{0})\Big]\Phi(r_s,\theta^0,\xi)\ , \label{eq:xi}
\end{eqnarray}
with $\theta^0= \theta (1+\xi)^{2/3}$ and the interpolation function 
\begin{align}
\Phi(r_s,\theta,\xi) &= \frac{ (1+\xi)^{\alpha(r_s,\theta)} + (1-\xi)^{\alpha(r_s,\theta)} -2 }{2^{\alpha(r_s,\theta)}-2} , \label{eq:phi}
\\
\alpha(r_s,\theta) &= 2 - h(r_s) e^{-\theta\lambda(r_s,\theta)}, 
\nonumber\\
h(r_s) &= \frac{2/3+h_1 r_s}{1+h_2 r_s}, \quad
\lambda(r_s,\theta) = \lambda_1 + \lambda_2 \theta r_s^{1/2}\ .
\nonumber
\end{align}
First, $h_1$ and $h_2$ are obtained by fitting $f_\text{xc}(r_s,0,\xi)$ to the ground state data of Ref.~\cite{spink} for $\xi=0.34$ and $\xi=0.66$. Subsequently, we use our extensive new QMC data set for $v^\xi(r_s,\theta)$ [107 data points for $\xi=1/3$ and $\xi=0.6$] to determine $\lambda_1$ and $\lambda_2$. 
Interestingly, we find that the spin interpolation depends only very weakly on $\theta$, i.e., $\lambda_2$ vanishes within the accuracy of the fit and, thus, we set $\lambda_2=0$. Remarkably, a single fit parameter ($\lambda_1$) is sufficient to capture the entire $\theta-$dependence of the spin interpolation function with a mean and maximum deviation from the QMC data at intermediate $\xi$ of $0.15\%$ and $0.8\%$.

Note that this is the first time that $\Phi(r_s,\theta,\xi)$ is obtained accurately from \textit{ab initio} data. Previous spin interpolations were based on uncontrolled approximations \cite{pdw_param,ksdt,tanaka_spin}.
A comparison of the $\xi$-dependence of $f_\text{xc}$ with various earlier parametrizations is depicted in Fig.~\ref{fig:xi}. The IIT and Tanaka curves utilize a different spin interpolation ~\cite{tanaka_spin} that is nonlinear in $f_\text{xc}^0$ and $f_\text{xc}^1$. These differences are most pronounced at intermediate temperature. The KSDT parametrization utilizes the functional form of $\Phi$ from Eq.~(\ref{eq:phi}). However, 
due to the absence of RPIMC data for intermediate $\xi$ they used the classical mapping data of Ref.~\cite{pdw_param} to determine the coefficients of $\Phi$. Overall, the KSDT fit is closest to our new parametrization at low $\theta$, while for $\theta>1$ the IIT curve is more accurate.
Nevertheless, we conclude that no previous model satisfactorily captures the correct $\xi$-dependence uncovered by our new \textit{ab initio} data.

{\bf Summary and discussion}.
In summary, we have presented the first \textit{ab initio} XC free energy functional of the UEG at WDM conditions, achieving an unprecedented accuracy of $\Delta f_\text{xc}/f_\text{xc}\sim0.3\%$. To cover the entire relevant parameter range, we carried out extensive {\em ab initio} QMC simulations for multiple spin polarizations at $0.1\leq r_s \leq 20$ and $0.5\leq\theta\leq 8$. 
In addition, we obtained accurate data for $0.0625\leq\theta\leq0.25$ by combining ground state QMC results with a small STLS temperature correction. 
All of our results are tabulated in the Supplemental Material~\cite{supplement} and  provide benchmarks for the development of new theories and simulation schemes as well as for the improvement of existing models.

The first step in our construction of the complete XC functional, $f_{xc}(r_s,\theta,\xi)$, was to obtain parametrizations for the completely polarized and unpolarized cases. This was achieved by fitting the RHS of Eq.~(\ref{eq:fit}) to our new data for the interaction energy $v^\xi$, for $\xi=0$ and $\xi=1$. The resulting parametrization reproduces the input data with a mean deviation of $0.17\%$ which is better by at least an order of magnitude compared to the KSDT fit. As an additional test of our parametrization, we performed independent QMC calculations of $e_\text{xc}$ (the XC energy per electron), for a wide range of values of $\theta$ down to $\theta=0.0625$,
and compared them to  $e_{xc}$-values calculated from our XC functional ($f_{xc}$). The striking agreement obtained constitutes strong evidence for the validity of Eq.~(\ref{eq:correction}) and consistency of our work.

Equipped with our new \textit{ab initio} XC-functional, we have also investigated the systematic errors of previous parametrizations. Overall, the functional by Ichimaru \emph{et al.}~\cite{tanaka_old, tanaka_new} deviates the least from our results, although at $\xi=1$ the classical mapping results by Perrot and Dharma-wardana~\cite{pdw_param} are similarly accurate. The KSDT parametrization exhibits large deviations exceeding $10\%$ for high temperature and density. At low temperatures, however, it performs surprisingly well, in part because it does not reproduce the systematic biases in the RPIMC data on which it was based. 

The construction of the first \textit{ab initio} spin interpolation function $\Phi(r_s,\theta,\xi)$ at WDM conditions constitutes the capstone of this work. Surprisingly, we find that a one-parameter fit is sufficient to capture the whole temperature dependence of the spin interpolation function. Further, we show that no previously suggested spin interpolation gives the correct $\xi$ dependence throughout the WDM regime.  

We are confident that our extensive QMC data set and accurate parametrization of the thermodynamic functions of the warm dense electron gas will be useful in many applications. Given recent developments in thermal Kohn-Sham DFT~\cite{shen1,shen2}, time-dependent Kohn-Sham DFT~\cite{tddft}, and orbital free DFT~\cite{oft1,oft2}, our parametrization of $f_\text{xc}$ is directly applicable for calculations in the local spin-density approximation. Furthermore, our functional can be used as a basis for gradient expansions~\cite{gga,gga2}, or as a benchmark for nonlocal functionals based on the fluctuation-dissipation theorem~\cite{burke2}.
In addition, it can be straightforwardly incorporated into widely used approximations in quantum hydrodynamics ~\cite{michta,manfredi} or for the equations of state of astrophysical objects~\cite{pot1,sauron1,sauron2}.
Finally, our XC functional should help resolve several exciting and controversial issues in warm dense matter physics, such as the existence and locations of the phase transitions in warm dense hydrogen~\cite{norman_starostin68, fortov_07, pierleoni16} or details of hydrogen-helium demixing~\cite{morales}.

Computational implementations of our XC functional (in FORTRAN, C++, and Python) are available online~\cite{our_git}.

\section*{Acknowledgements}
This work was supported by the Deutsche Forschungsgemeinschaft via project BO1366-10 and via SFB TR-24 project A9 as well as grant shp00015 for CPU time at the Norddeutscher Verbund f\"ur Hoch- und H\"ochstleistungsrechnen (HLRN).
TS~acknowledges the support of the US DOE/NNSA under Contract No.~DE-AC52-06NA25396.
FDM is funded by an Imperial College PhD Scholarship.
FDM and WMCF used computing facilities provided by the High Performance Computing Service of Imperial College London, by the Swiss National Supercomputing Centre (CSCS) under project ID s523, and by ARCHER, the UK National Supercomputing Service, under EPSRC grant EP/K038141/1 and via a RAP award.
FDM and WMCF acknowledge the research environment provided by the Thomas Young Centre under Grant No.~TYC-101.


\begin{thebibliography}{26}
\bibitem{ksdt-bcdc-note}
In contrast, the KSDT interaction energy (solid blue line) exhibits severe deviations to the corresponding BCDC data (blue dots). This indicates that using the systematically biased RPIMC data as input for a parametrization of $f_\text{xc}$ does not allow to consistently recover other quantities (see also the discussion of Fig.~\ref{fig:best_plot_in_human_history}).
\bibitem{ksdt-note-2}
Note that the KSDT parametrization is the result of a direct fit to the RPIMC data for $e_\text{xc}$ but only reproduces these data for $\theta\geq1$ and strongly deviates below (the KSDT fit exhibits a mean and maximum deviation to the underlying RPIMC data of $1.2\%$ and $7.8\%$ for $\xi=1$).
\end{thebibliography}
\end{document}